\documentclass[journal,twoside,web]{ieeecolor}
\usepackage{tmi}
\usepackage{cite}
\usepackage{amsmath,amssymb,amsfonts}
\usepackage{algorithmic}
\usepackage{graphicx}
\usepackage{textcomp}
\usepackage{cite}
\usepackage{amsmath,amssymb,amsfonts}
\usepackage{bbold}
\usepackage{graphicx}
\usepackage{textcomp}
\usepackage{soul, color}
\usepackage{steinmetz}
\usepackage{comment}
\usepackage{algorithm2e}
\SetKwComment{Comment}{/* }{ */}
\SetKwInput{kwPreparation}{Preprocessing}
\SetKwInput{kwinput}{Input}
\SetKwInput{kwoutput}{Output}
\SetKwInput{kwPreparation}{Preprocessing}
\SetKwInput{kwTraining}{Training}
\SetKwInput{kwInference}{Evaluation}
\RestyleAlgo{ruled}
\def\BibTeX{{\rm B\kern-.05em{\sc i\kern-.025em b}\kern-.08em
    T\kern-.1667em\lower.7ex\hbox{E}\kern-.125emX}}
\begin{document}
\title{The Art of the Steal: Purloining Deep Learning Models Developed for an Ultrasound Scanner to a Competitor Machine}
\author{Ufuk Soylu, \textit{Student Member, IEEE}, Varun Chandrasekeran \textit{Member, IEEE} and Michael L. Oelze, \textit{Senior Member, IEEE}
\thanks{This research received financial support from grants provided by the National Institutes of Health (NIH) (R01CA251939 and R01CA273700). Submitted for review on XX.}
\thanks{Ufuk Soylu, Varun Chandrasekaran and Michael Oelze are with the Department of Electrical and Computer Engineering, University of Illinois at Urbana-Champaign, Urbana, IL 61801 USA. Moreover, Michael Oelze is with the Beckman Institute for Advanced Science and Technology and the Carle Illinois College of Medicine, University of Illinois at Urbana-Champaign, Urbana, IL 61801 USA. (e-mail: usoylu2@illinois.edu; varunc@illinois.edu, oelze@illinois.edu).}
}

\maketitle
\begin{abstract}
A transfer function approach has recently proven effective for calibrating deep learning (DL) algorithms in quantitative ultrasound (QUS), addressing data shifts at both the acquisition and machine levels. Expanding on this approach, we develop a strategy to 'steal' the functionality of a DL model from one ultrasound machine and implement it on another, in the context of QUS. This demonstrates the ease with which the functionality of a DL model can be transferred between machines, highlighting the security risks associated with deploying such models in a commercial scanner for clinical use. The proposed method is a black-box unsupervised domain adaptation technique that integrates the transfer function approach with an iterative schema. It does not utilize any information related to model internals of the victim machine but it solely relies on the availability of input-output interface. Additionally, we assume the availability of unlabelled data from the testing machine, i.e., the perpetrator machine. This scenario could become commonplace as companies begin deploying their DL functionalities for clinical use. Competing companies might acquire the victim machine and, through the input-output interface, replicate the functionality onto their own machines. In the experiments, we used a SonixOne and a Verasonics machine. The victim model was trained on SonixOne data, and its functionality was then transferred to the Verasonics machine. The proposed method successfully transferred the functionality to the Verasonics machine, achieving a remarkable 98\% classification accuracy in a binary decision task. This study underscores the need to establish security measures prior to deploying DL models in clinical settings.
\end{abstract}
\section{Introduction}
\label{sec:intro}

The integration of DL-based biomedical ultrasound has the potential to enhance the quality of medical services through automation and efficiency. Therefore, the integration of DL-based biomedical ultrasound imaging into clinical practice has become the coveted goal for both industry and academia. Thanks to substantial efforts dedicated to overcoming challenges posed by DL-based models in clinical settings, this goal now can be within reach. The primary technical challenges stem from data scarcity, attributed to the high costs involved in conducting laboratory experiments, acquiring human image data, and acquiring expert annotations of data. Further exacerbating this is the presence of data mismatch, arising from inevitable discrepancies between the development and clinical environments \cite{castro2020causality}. These challenges have been well studied and several methods have been proposed to overcome these difficulties. Although ongoing efforts persist in addressing these challenges, DL-powered algorithms have reached a stage of maturity and increasingly adoption of such algorithms will occur clinically.

The field of Quantitative Ultrasound (QUS) has also shifted from traditional methods to DL-based approaches, following the general trend. Specifically, in several recent examples, DL models have demonstrated superior performance to  traditional QUS approaches in classifying tissue states \cite{byra2022joint, byra2022liver,han2020noninvasive, nguyen2019reference, nguyen2021use, taleghamar2022deep, jeon2023two}. Accordingly, substantial efforts dedicated to overcome challenges related to clinical adoption, namely data scarcity and data mismatches have been investigated \cite{tehrani2022robust, kim2021learning,tierney2020domain, tehrani2021ultrasound, soylu2023calibrating, sharifzadeh2021ultrasound, soylu2022data, soylu2023machine}. Observing the pace of the development in this field, it is reasonable to claim that DL-based QUS approaches in clinical setting are becoming a reality rather than a distant goal.

Given the advancement, there could still be concerns or undiscovered challenges on the path that can delay the wide adoption of DL-based QUS methods in clinic. In this work, we identify one such concern: the ease with which the functionality of DL-based QUS methods can be stolen from one machine and implemented on another, competitor machine. There can be great expense associated with the development of DL models for medical diagnostics. Specifically, a major expense is in the accumulation of large amounts of data with expert annotation and labelling. The cost of labelling copious amounts of data for DL training can be expensive and take years of investment. Companies investing in the development of such functionalities may be deterred from deploying their DL-based models to their own machine and making it available for clinical use, if competitors can effortlessly access their machine and transfer their functionality. Therefore, as an outcome of this work, we highlight the necessity of improving DL-model security in the context of biomedical ultrasound imaging by demonstrating the ability to purloin a DL-model from one machine, i.e., the victim machine, and deploy the DL-model on a different machine, i.e., the perpetrator machine.

The most relevant framework to this study is black-box unsupervised domain adaptation (UDA). In black-box UDA, we don't have access to the model's internals which means that its weights, its architecture, training process, and training data are unknown. However, we have access to input-output interface and there is unlabeled data available for the machine to which we aim to transfer functionality. This is the most relevant framework for security threats in deployment of DL-based models in the clinic. From the perspective of competitors, the process of stealing the model can be accomplished through a simple stepwise process. First, the competitor acquires the victim machine, gaining access to its input-output interface. Specifically, image data generated from the perpetrator machine could be transferred to the hard drive of the victim machine assuming the file structure can be replicated so that the victim machine reads the image data. Next, the competitor acquires unlabeled data utilizing their own machine in the pursuit of their own model development, making unlabeled data abundant. The unlabelled data or images from the perpetrator machine are transferred to the victim machine, i.e., test-time calibration \cite{soylu2023machine}. Labels are given to each image from the DL model on the victim machine. These labels are then used in the perpetrator machine to train the model.  

A comprehensive literature survey on adaptation algorithms at test-time can be found in \cite{liang2023comprehensive}. Black-box UDA falls under the umbrella of test-time adaptation. There exist multiple strategies to address the black-box UDA problem: pseudolabeling aims to assign labels for unlabeled data via the black-box model. Consistency regularization aims to enforce consistent network predictions by adding a loss term in the training. Clustering assumes that the decision boundary in the unlabeled data should lie in low-density regions. Self-supervised learning aims to learn feature representation from unlabeled data based on auxiliary prediction task to be used in the down-streaming task. Given the vast literature, IterLNL \cite{zhang2021unsupervised} is a state-of-the-art black-box UDA method which conducts noisy labeling and learning with noisy labels, iteratively. Combining IterLNL with our transfer function approach \cite{soylu2023calibrating, soylu2023machine} led to successful transfer of the DL functionality from the victim machine to the perpetrator machine.

\begin{figure}[hbt!]
\begingroup
    \centering
    \begin{tabular}{c}
\hspace{-0.5cm}{ \includegraphics[width=1.03\linewidth]{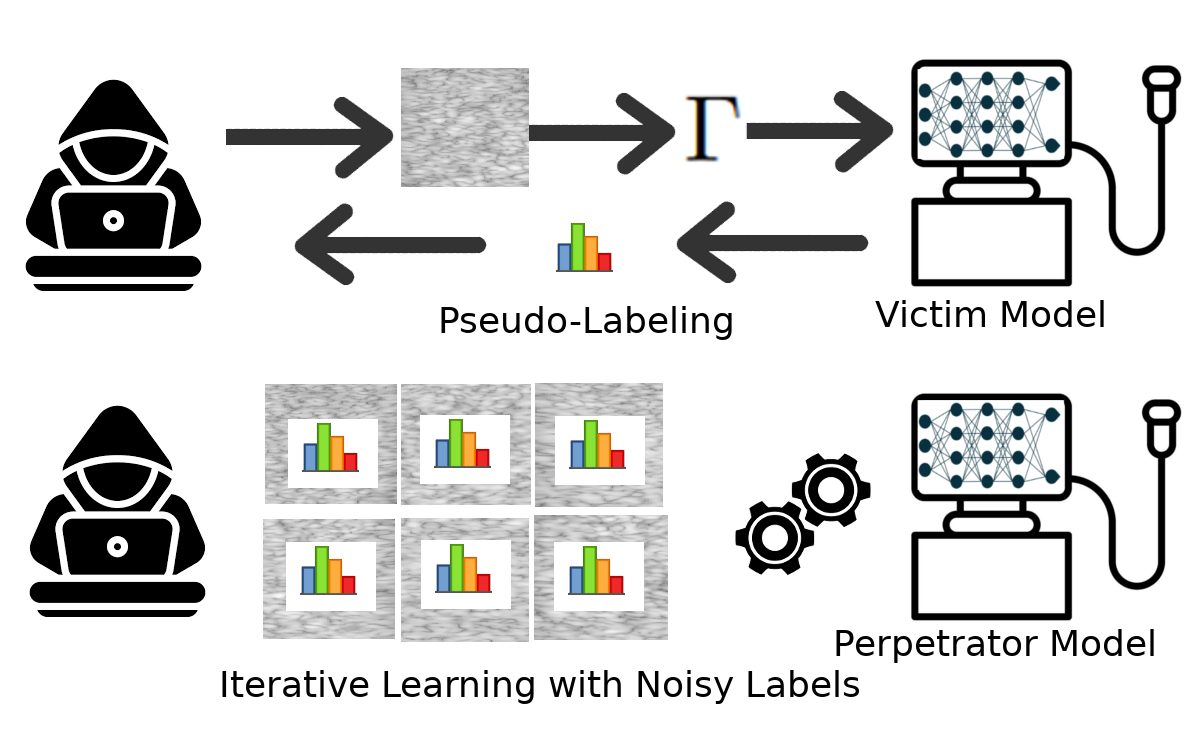}}\\
\end{tabular}
\caption{The proposed method}
\label{fig:proposed}
\endgroup
\end{figure}

The proposed method, depicted in Fig. \ref{fig:proposed}, utilizes a transfer function at test-time to calibrate mismatches between the victim and the perpetrator machines. Subsequently, pseudolabels can be obtained for unlabeled data from the perpetrator machine through the input-output interface of the victim machine. These labels can still be noisy. Therefore, we propose utilizing IterLNL to further refine the labels. Then, a new DL model can be trained for the perpetrator machine utilizing these refined labels, which copies the functionality of DL-based model from the victim to the perpetrator machine. The proposed method accesses only the input-output interface without accessing model internals of the victim machine, which are presumed not to be available. Another note is that the transfer function method requires the acquisition of calibration data using the victim and the perpetrator machines with a calibration phantom. Further details of the proposed approach will be discussed in Section \ref{sec:approach}. In Sections \ref{sec:met} and \ref{sec:results}, our methodology and experimental results are detailed, respectively. Following that, we provide discussion and conclusions in Section \ref{sec:discussion} and in Section \ref{sec:conclusion}, respectively.

\section{Proposed Approach}
\label{sec:approach}

The proposed approach combines the transfer function \cite{soylu2023calibrating, soylu2023machine} with IterLNL \cite{zhang2021unsupervised}. The transfer function is able to calibrate data mismatches between ultrasound machines over a defined bandwidth. It has been demonstrated that utilizing the transfer function between two machines at test-time, i.e., converting data acquired on one machine to the DL model developed on another machine, achieved approximately an AUC scores of 0.99 and accuracy of 80 percent for a QUS task. Therefore, as a refinement step, IterLNL, which conducts iterative learning with noisy labels, is performed to further refine the labels. Following that, a new learning algorithm can be developed using these refined labels for the perpetrator machine. Therefore, the proposed approach at a high level involves 5 steps:
\begin{itemize}
    \item[1.] Gather calibration data using the victim machine and the perpetrator machine.
    \item[2.] Calculate the transfer function between the perpetrator and the victim machines.
    \item[3.] Obtain pseudo-labels for the unlabeled data from the perpetrator machine utilizing the transfer function and the input-output interface of the victim machine.
    \item[4.] Implement iterative learning with pseudo-labels for refinement.
    \item[5.] Obtain the learning algorithm for the perpetrator machine using the refined pseudo-labels.
\end{itemize}
The transfer function approach at test-time exploits the decomposition of an ultrasound image's frequency spectrum into the tissue signal and the system response,
\begin{align}
\label{powerspec_simple}
    I_{ultrasound}(\mathsf{f}, \mathsf{x}) = S_{\phi_{machine}}(\mathsf{f},\mathsf{x}) P_{tissue}(\mathsf{f},\mathsf{x})
\end{align}
where $\mathsf{x}$ is the axial location, $\mathsf{f}$ is the frequency $S_{\phi_{machine}}$ is the system response and  $P_{tissue}$ is the imaging  substrate. Then, the transfer function can be calculated using a calibration phantom and obtaining a single view from a calibration phantom utilizing both machines,  
\begin{align}
\label{powerspec}
   \frac{I_{victim}(\mathsf{f}, \mathsf{x})}{I_{perpetrator}(\mathsf{f}, \mathsf{x})} &= \frac{S_{\phi_{victim}}(\mathsf{f}, \mathsf{x})}{S_{\phi_{perpetrator}}(\mathsf{f}, \mathsf{x})}\\ &= \Gamma
\end{align}
where $\Gamma$ is the transfer function at test-time.  Following the methodology of the original transfer function work, we computed the transfer function at various depths and employed the Wiener implementation of it,
\begin{align}
\label{wiener train}
    \Gamma^{Wiener} &= \frac{|\Gamma|^{-1}}{|\Gamma|^{-2}+SNR^{-1}}.
\end{align}

For simplicity, $\Gamma^{Wiener}$ will be referred as $\Gamma$ for the rest of the paper. Actually, we obtained the transfer function in two ways: stable acquisition, which involved capturing 10 calibration views from a fixed location of the calibration phantom, and free-hand acquisition, which involved recording a video of calibration frames using free-hand motion. While stable acquisition is the most standard way to obtain calibration data, free-hand acquisition is necessary in scenarios where stable acquisition is not possible. In this study, in one of the experiments, we utilized different transducers for the victim and perpetrator machines and had to implement free-hand acquisition for the calibration data.

The key idea in IterLNL is that given noisy labels, a new model is learned iteratively based on anchors, which are the data points exhibiting the lowest training loss. Let's denote the victim model $F_{victim}$, unlabeled data $X_{perpetrator}$, the perpetrator model $F_{perpetrator}$ and anchors $U_{perpetrator}$. Then, pseudo-labels $y_{perpetrator}$ can be obtained by applying the input-output interface of $F_{victim}$ to $ \Gamma (X_{perpetrator}) = X_{perpetrator \rightarrow victim}$. At each iteration, pseudo-labels will be acquired again using the learnt perpetrator model $F_{perpetrator}$. An important parameter in IterLNL is the noise rate, denoted as $\epsilon$. In the original work, it was estimated using $y_{perpetrator}$ and $X_{perpetrator \rightarrow victim}$. However, in this study, we leveraged our prior knowledge regarding the transfer function, establishing the noise rate at approximately 20 percent. Then, the proposed approach that combines the transfer function and IterLNL can be depicted in Algorithm \ref{alg:algo_iter_test}. One note is that after the iterative learning process, $F_{perpetrator}$ can generate highly accurate labels for the unlabeled data $X_{perpetrator \rightarrow victim}$, which has been transformed using the transfer function $\Gamma$. As a final step, we may update $F_{perpetrator}$ so that it operates directly within the perpetrator machine domain. This involves training it directly on $X_{perpetrator}$ without the application of the transfer function $\Gamma$. 

\begin{algorithm}[hbt!]
\caption{Proposed Approach}\label{alg:algo_iter_test}
\kwinput{Victim Model $F_{victim}$, Unlabeled data $X_{perpetrator}$, Transfer Function $\Gamma$, $\epsilon = 20$}
\kwoutput{Perpetrator Model $F_{perpetrator}$}
Step1: Apply Transfer Function $X_{perpetrator \rightarrow victim}$ $\xleftarrow{}$ $\Gamma(X_{perpetrator})$ \;
Step2: Standardization $X_{perpetrator \rightarrow victim} \xleftarrow{} (X - \mathbb{E}X_{perpetrator \rightarrow victim})/\sigma X_{perpetrator \rightarrow victim}$\;
Step3: Set $F = F_{victim}$ as the pseudo-labeling model\;
\kwTraining{Iterative Learning}
\While{iteration}{
Acquire noisy labels $y_{perpetrator}$ using $F$\;
Initialize $F_{perpetrator}$\;
Obtain anchors $U_{perpetrator}$ (100-$\epsilon$ percent of $X_{perpetrator \rightarrow victim}$ returning lowest $Loss(F_{perpetrator}(X_{perpetrator \rightarrow victim}), y_{perpetrator})$\;
\While{epoch}{
Update $F_{perpetrator}$ using anchors $U_{perpetrator}$\;
Update anchors $U_{perpetrator}$\;
}
Update $F$ as $F_{perpetrator}$}
Final Step: Obtain refined pseudo-labels $y_{perpetrator} = F(X_{perpetrator \rightarrow victim})$ and retrain $F_{perpetrator}$ using $(X_{perpetrator}, y_{perpetrator})$ which works directly on the perpetrator machine without $\Gamma$
\end{algorithm}

\section{Methods}
\label{sec:met}

\subsection{Phantoms}

\begin{figure}[hbt!]
\begingroup
    \centering
    \begin{tabular}{c c c}
\hspace{-0.25cm}{ \includegraphics[scale=0.03]{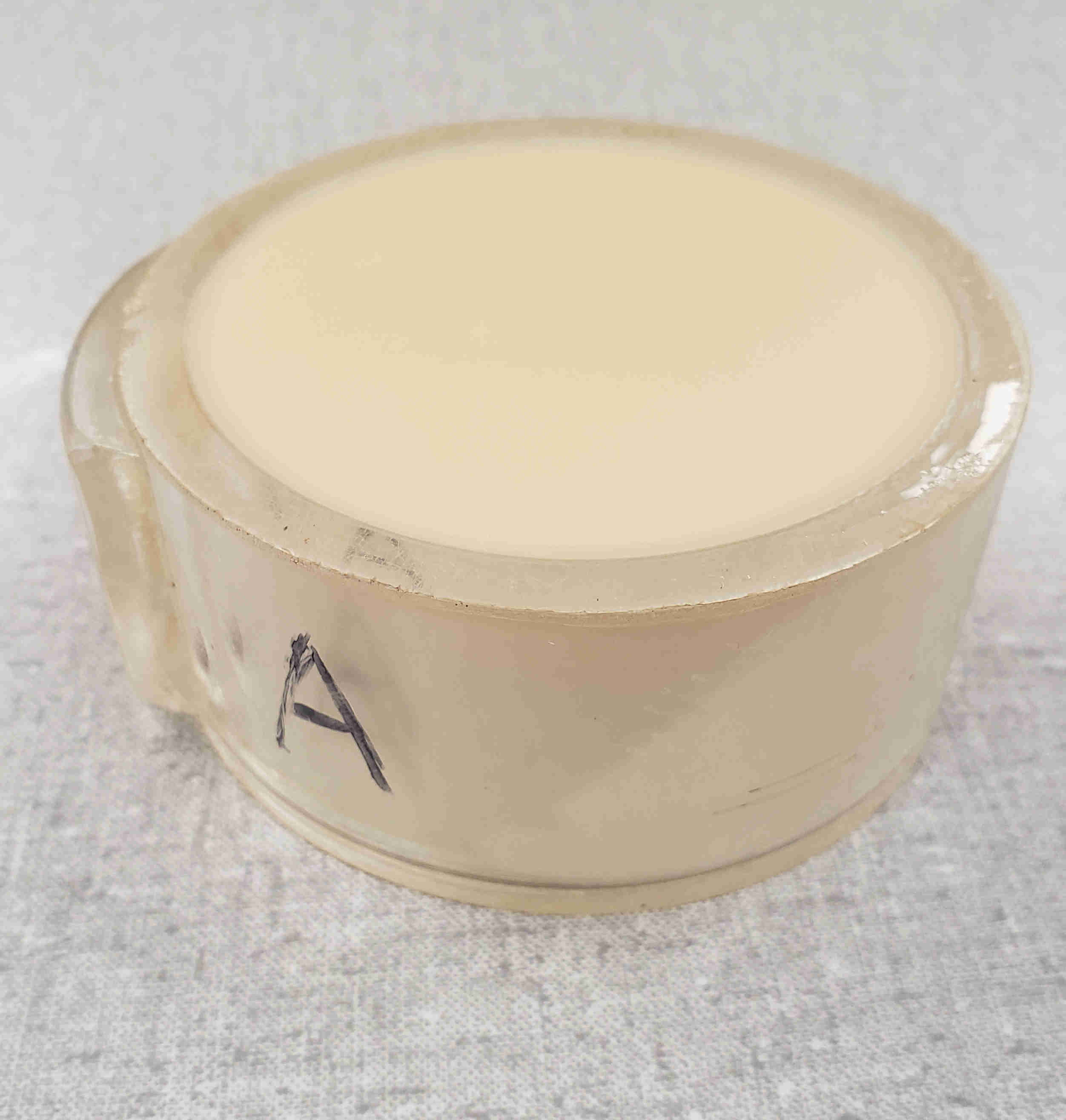}}&\hspace{-0.25cm}{ \includegraphics[scale=0.029]{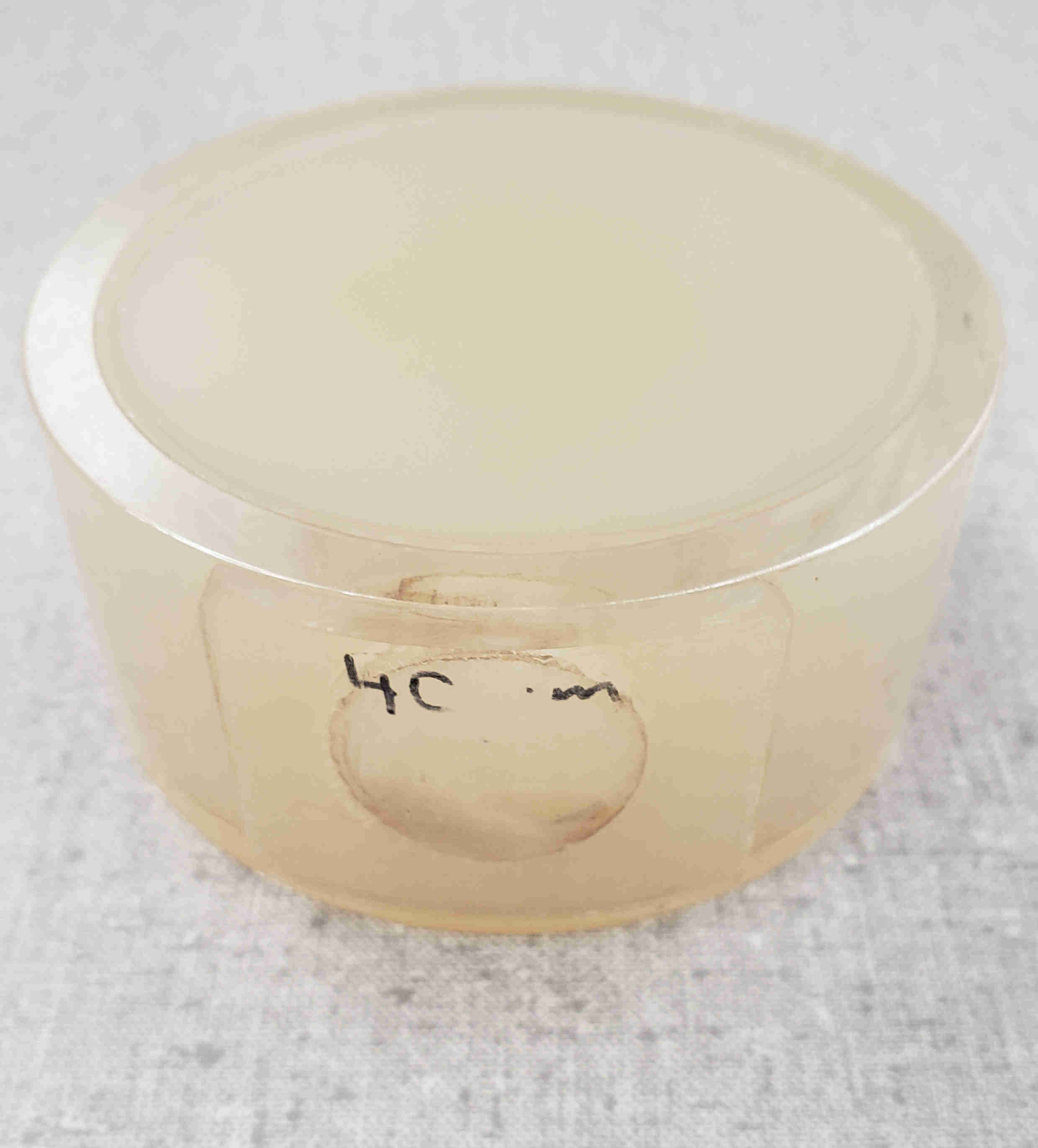}} \hspace{-0.cm}{ \includegraphics[scale=0.0264]{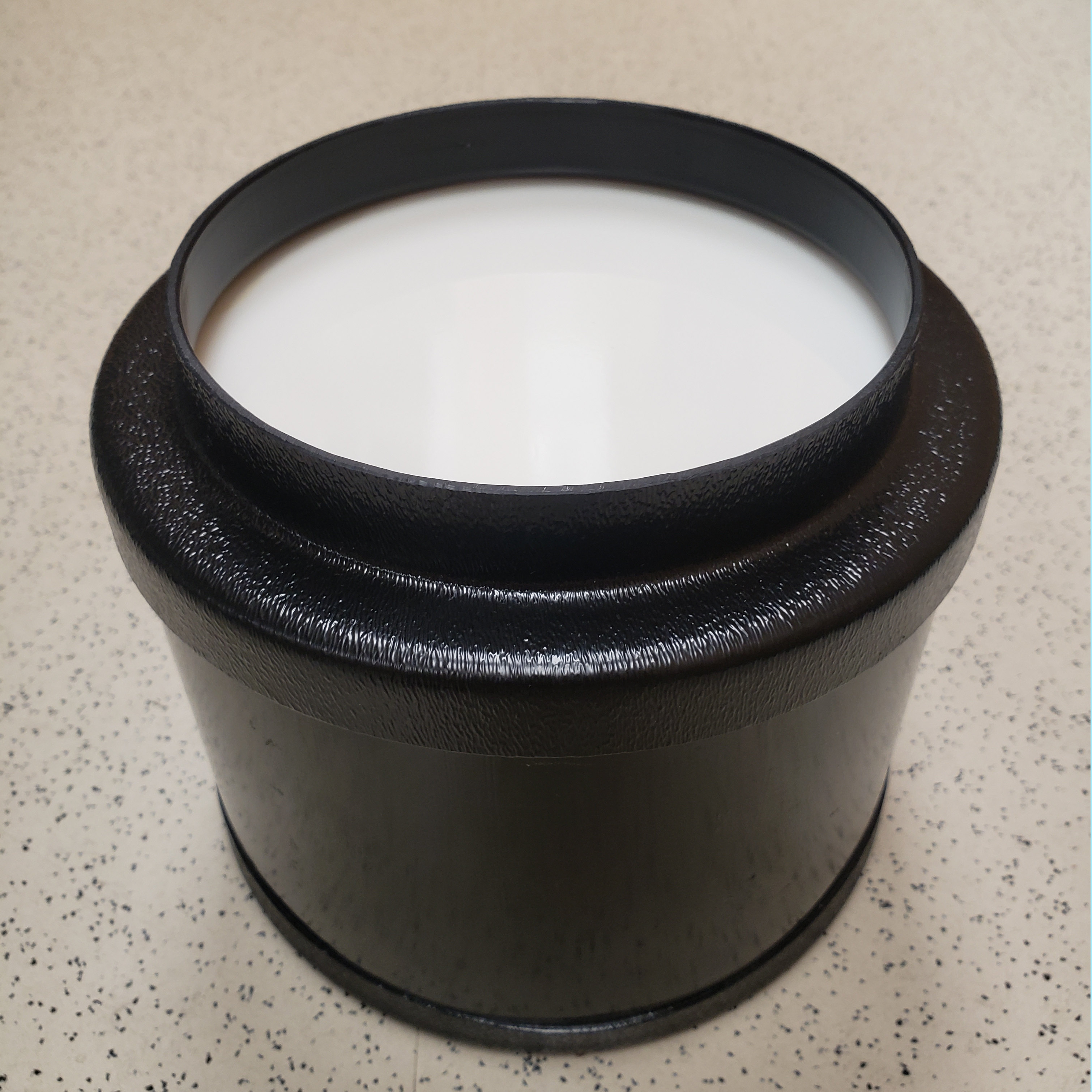}} \\
\end{tabular}
\caption{Classification Phantoms and  Calibration Phantom}
\label{fig:phantom_picutes}
\endgroup
\end{figure}

The experiments used one calibration phantom and two classification phantoms with their photographs shown in Fig. \ref{fig:phantom_picutes}. The QUS task in this study was to determine the class identity of the ultrasound data from each phantom. Clinically, this could correspond to a binary tissue classification problem, e.g., is the tissue diseased or not diseased, or is the the tumor malignant or benign, or is the liver fatty or normal. However, in this study we limit the application to classifying between two phantoms to validate the approach. The objective of the study is to steal this functionality from the victim machine and implement it on the perpetrator machine. 

A commercial QUS reference phantom from CIRS, Inc., Norfolk, VA (part no. 14090502, serial no. 221447541) was used as the calibration phantom. The phantom had a speed of sound (SOS) of 1545 meters per second and an attenuation coefficient slope (ACS) of 0.74 dB per centimeter per megahertz.

Classification Phantom 1 was designed to mimic the characteristics of the human liver \cite{wear2005interlaboratory}, with construction details provided in \cite{madsen1998liquid}. It contained glass-bead scatterers. The beads had sizes in a range between 75 and 90 $\mu$m in diameters. The phantom had an ACS of 0.4 dB per centimeter per megahertz, with a SOS of 1540 meters per second.

Classification Phantom 2 was described as low attenuation characteristics\cite{anderson2010interlaboratory},  with construction details provided in \cite{madsen1978tissue}. It contained glass-bead scatterers. The beads had sizes in a range between 39 and 43 $\mu$m in diameters. The phantom had an ACS of 0.1 dB per centimeter per megahertz, with a SOS of 1539 meters per second.

\vspace{1cm}
\subsection{Ultrasound Machines}

Two ultrasound machines, a SonixOne system and a Verasonics system, were utilized. The victim model was trained using data from the SonixOne system, which designated it as the victim machine. The perpetrator model aimed to replicate the same functionality as the victim model but for the Verasonics system, positioning it as the perpetrator machine. Throughout the experiments, an L9-4 transducer and an L11-5 transducer were utilized. The L9-4 transducer was used for the SonixOne system, while both the L9-4 and L11-5 transducers were used for the Verasonics system. 

To prepare the perpetrator image data from the Verasonics system for the input-output interface on the victim machine , it was necessary to adjust the sampling rate mismatch between the two machines. While the Verasonics machine captured raw channel data at a 50 MHz sampling rate, the SonixOne machine obtained radio-frequency (RF) data at a 40 MHz sampling rate after it had been beamformed. These differences were mitigated by preprocessing the Verasonics data. First, delay-and-sum beamforming was applied to the raw channel data. Next, a multirate finite impulse response filter with an upsampling factor of 4 and a downsampling factor of 5 was used to correct the sampling rate mismatch. All these preprocessing operations were done on Matlab, specifically using the 'designMultirateFIR' and the 'dsp.FIRRateConverter' functions. This way, the Verasonics data was matched to the SonixOne data in terms of sampling rate.
The victim model was trained using 1,000 frames per classification phantom, totaling 2,000 frames, acquired using the SonixOne machine. Similarly, for the perpetrator model, 2,000 frames were acquired using the Verasonics machine. These acquisitions involved recording a video of frames while moving the transducer across the phantom surface. 

For calibration data, both the SonixOne and Verasonics machines were utilized. By securing the transducer using a bar clamp holder, ten identical frames were captured from precisely the same position on the calibration phantom. Multiple frames were recorded to mitigate or reduce any potential effects of electrical noise through averaging.

Line-by-line scanning was conducted for both machines. During acquisition, a single axial focus was set at 2 cm. The Verasonics system operated at a pulse frequency of 5 MHz and an output power of 45.2 Volts. In contrast, the SonixOne system operated at a pulse frequency of 9 MHz and an output power of 0 dB.

\subsection{Data Preparation}

An ultrasound image frame from both machines was 2080 pixels in height and 256 pixels in width, corresponding to 4 cm axial depth and 128 array channels. The victim and the perpetrator model used the RF ultrasound data. From the ultrasound images, rectangular patches were obtained for training and evaluating the DL models. They had a height of 200 samples and a width of 26 samples, equivalent to 4 mm by 4 mm. In traditional QUS, a region of interest in the ultrasound frame is extracted and analyzed. The patch-wise data processing in DL models in this work was inspired by the nature of traditional QUS.

\begin{figure}[hbt!]
\begingroup
    \centering
    \begin{tabular}{c}
\hspace{-0.0cm}{ \includegraphics[width=0.7\linewidth]{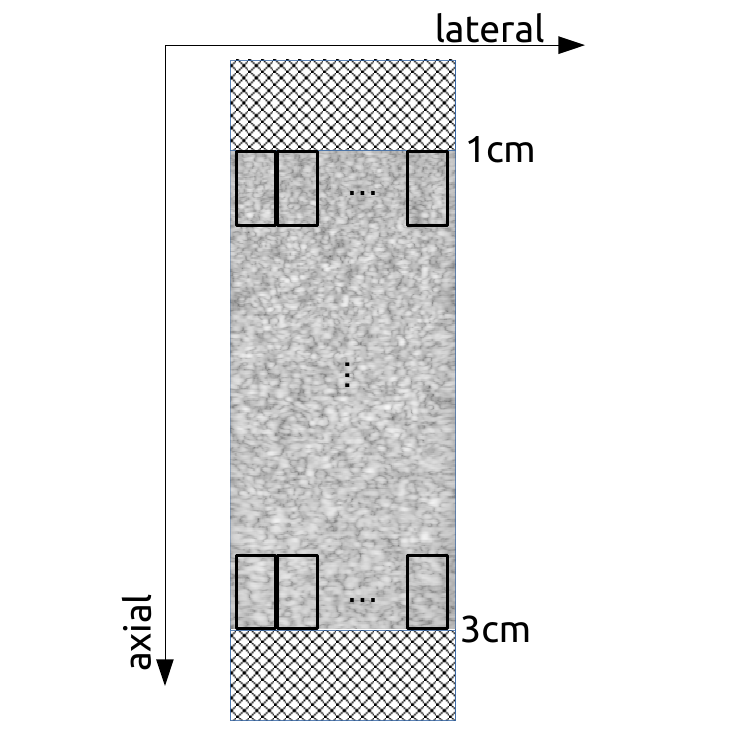}}\\
\end{tabular}
\caption{QUS Data Patches}
\label{fig:patch_extract}
\endgroup
\end{figure}

The patch extraction is depicted in Fig. \ref{fig:patch_extract}. The initial 540 axial pixels were skipped. Axially, the next sequence started after shifting 100 pixels. Laterally, the next sequence started after shifting 26 pixels. Overall, 81 patches per ultrasound frames were extracted coming from 9 lateral and 9 axial locations.

The 2,000 frames from the SonixOne machine were split into a training and a validation set with a ratio of 4:1 for the development of the victim model. On the other hand, for the development of the perpetrator model, 2000 frames from the Verasonics machine were split into a training and a testing set with a ratio of 1:1. Therefore, in the experiments, the perpetrator model was trained using the training set consisting of 1000 frames, and it was evaluated using the test set consisting of 1000 ultrasound frames.

\subsection{Training}

During training, the Adam algorithm \cite{kingma2014adam} was used as the optimizer. The cross-entropy loss was used for training. Z-score standardization was conducted patch-wise, which involved subtracting the patch-wise mean and normalizing it with the patch-wise standard deviation. A batch size of 2048 was employed for all experiments to maximize memory utilization. Data augmentation involved horizontally flipping the patches with a probability of half. 

In iterative learning, during each iteration, ten percent of the data was held out for use as a validation partition. This allocation allowed for the observation of training metrics and the determination of hyperparameters, such as setting the epoch number to 10 and the learning rate to $1\times10^{-5}$. In results, the experiments were repeated ten times starting from the iterative learning. The performance of the final $F_{perpetrator}$ on the hidden data from the Verasonics machine was reported. The performance evaluation metrics included average classification accuracy and AUC score, along with their respective standard deviations, calculated patch-wise. The observed standard deviation was due to random parameter initialization in each iteration.

For computational resources, four NVIDIA RTX A4000 GPUs were used in parallel in a workstation. The PyTorch library \cite{paszke2019pytorch} was utilized for the implementation.

\subsection{Network Structure}

The victim model was a DenseNet-201 \cite{huang2017densely} with minor tweaks to their input-output relationships tailored to our specific problem. The perpetrator model was chosen as ResNet-34 \cite{he2016deep} with similar tweaks. These tweaks involve adjusting the first convolutional layer to accommodate single-layer inputs and modifying the final output layer to output a single score for a binary classification setup. During the initialization of network parameters, pretrained weights were utilized, with the exception of the first 2D convolution layer and the last linear layer. They were randomly initialized utilizing PyTorch's default method. Then, all parameters were unfrozen and updated using backpropagation during training. The necessity of selecting different model architectures between the victim and the perpetrator arose from the black-box nature of the victim setup, resulting in inevitable differences in architectural design. As long as the selected architecture has enough complexity to solve the task, the attack should be successful. In practice, network selection may require search and hyperparameter tuning, especially in a black-box setup. However, if the perpetrator knows the complexity of the victim model or its architecture, the perpetrator can use the victim's architecture directly.

\section{Results}
\label{sec:results}

The victim model scored an accuracy of 99.93 percent and an AUC of 0.999 on its validation set. This performance serves as the upper bound achievable with the proposed method, acknowledging an anticipated performance degradation due to machine-level mismatches between the victim and perpetrator machines.

Our investigation primarily focused on four aspects. First, we explored the impact of the transfer function in our proposed approach. To examine this, we compared two cases: utilizing the transfer function and omitting it, essentially skipping step 1 in Algorithm \ref{alg:algo_iter_test}.  Second, our exploration delved into the importance of the priors utilized in the proposed method. These priors covered both the label distribution and the noise rate parameter $\epsilon$. Third, we explored the number of ultrasound frames from the perpetrator machine needed for successful attack. Fourth, we explored the effect of different transducers between the victim and the perpetrator machines. While in the first three experiments, only the L9-4 transducer was utilized, in the last experiment, the L9-4 transducer was used for the SonixOne system, and the L11-5 transducer was used for the Verasonics system.

\subsection{The Effect of the Transfer Function}
\label{sec:tf}
We conducted an ablation study for the proposed method consisting of two experiments. The first experiment involved using the transfer function within the iterative learning schema. In the second experiment, we omitted the transfer function and solely implemented the iterative learning schema. These experiments were designed to gauge the contribution of the transfer function method. The results for different iteration numbers (2, 5, and 10) are presented in Table \ref{tab:res1}.

\begin{table*}[t]
    \centering
\caption{Transfer Function}
{\renewcommand{\arraystretch}{1.2}
{\small
\begin{tabular}{ |p{0.25cm}||p{3cm}||p{1.8cm}||p{1.7cm}||p{1.8cm}||p{1.7cm}||p{1.8cm}||p{1.7cm}|}
 \hline
\textit{No}&\textit{Experiment}&\textit{Accuracy (Iter=2)}&\textit{AUC (Iter=2)}&\textit{Accuracy (Iter=5)}&\textit{AUC (Iter=5)}&\textit{Accuracy (Iter=10)}&\textit{AUC (Iter=10)}\\
 \hline
1&Iterative Learning with Transfer Function & \textbf{97.83$\pm$0.28} & \textbf{0.9976$\pm$5e-4} & \textbf{97.78$\pm$0.25} & \textbf{0.9975$\pm$5e-4}&\textbf{97.69$\pm$0.36}&\textbf{0.9974$\pm$7e-4}\\
 \hline
2&Iterative Learning without Transfer Function & 89.98$\pm$3.88 & 0.9618$\pm$3e-2 & 89.22$\pm$7.39 & 0.9401$\pm$6e-2&85.51$\pm$1.60&0.9020$\pm$1e-1\\
 \hline
\end{tabular}
}}
\label{tab:res1}
\end{table*}

\begin{table*}[t]
    \centering
\caption{Label Distribution}
{\renewcommand{\arraystretch}{1.2}
{\small
\begin{tabular}{ |p{0.25cm}||p{3cm}||p{1.8cm}||p{1.7cm}||p{1.8cm}||p{1.7cm}||p{1.8cm}||p{1.7cm}|}
 \hline
\textit{No}&\textit{Experiment}&\textit{Accuracy (Iter=2)}&\textit{AUC (Iter=2)}&\textit{Accuracy (Iter=5)}&\textit{AUC (Iter=5)}&\textit{Accuracy (Iter=10)}&\textit{AUC (Iter=10)}\\
 \hline
1&$40^{th}$ percentile & 97.35$\pm$1.44 & 0.9975$\pm$7e-4 & 97.71$\pm$0.33 & 0.9974$\pm$6e-4&97.70$\pm$0.51&0.9975$\pm$7e-4\\
 \hline
2&$50^{th}$ percentile & \textbf{97.83$\pm$0.28} & \textbf{0.9976$\pm$5e-4} & \textbf{97.78$\pm$0.25} & \textbf{0.9975$\pm$5e-4}&97.69$\pm$0.36&0.9974$\pm$7e-4\\
 \hline
3&$60^{th}$ percentile & 97.55$\pm$0.47 & 0.9974$\pm$5e-4 & 97.48$\pm$0.46 & 0.9972$\pm$7-e4&\textbf{97.82$\pm$0.38}&\textbf{0.9976$\pm$7e-4}\\
 \hline
\end{tabular}
}}
\label{tab:res2}
\end{table*}

\begin{table*}[t]
    \centering
\caption{Noise Rate $\epsilon$}
{\renewcommand{\arraystretch}{1.2}
{\small
\begin{tabular}{ |p{0.25cm}||p{3cm}||p{1.8cm}||p{1.7cm}||p{1.8cm}||p{1.7cm}||p{1.8cm}||p{1.7cm}|}
 \hline
\textit{No}&\textit{Experiment}&\textit{Accuracy (Iter=2)}&\textit{AUC (Iter=2)}&\textit{Accuracy (Iter=5)}&\textit{AUC (Iter=5)}&\textit{Accuracy (Iter=10)}&\textit{AUC (Iter=10)}\\
 \hline
1& 10 percent & \textbf{98.25$\pm$0.29} & \textbf{0.9983$\pm$5e-4} & \textbf{98.13$\pm$0.28} & \textbf{0.9982$\pm$4e-4}&\textbf{98.25$\pm$0.28}&\textbf{0.9983$\pm$5e-4}\\
 \hline
2& 20 percent& 97.83$\pm$0.28 & 0.9976$\pm$5e-4 & 97.78$\pm$0.25 & 0.9975$\pm$5e-4&97.69$\pm$0.36&0.9974$\pm$7e-4\\
 \hline
3& 30 percent & 97.60$\pm$0.82 & 0.9969$\pm$2e-3 & 97.02$\pm$1.57 & 0.9954$\pm$4e-3&96.45$\pm$2.46&0.9932$\pm$1e-2\\
 \hline
\end{tabular}
}}
\label{tab:res3}
\end{table*}

\subsection{Priors}
\label{sec:prior}
We conducted a robustness study for the proposed method consisting of two sets of experiments. The proposed method utilized two information priors: label distribution and noise rate $\epsilon$. We utilized the prior for the label distribution to obtain noisy labels through the pseudo-labeling model. In our balanced binary classification setup, where half of the unlabeled data was expected to belong to one class and the other half to the other, we used this prior knowledge when assigning class identities in pseudo-labeling. Scores were thus thresholded at the median score within the training set. In Table \ref{tab:res2}, we explored scenarios where the threshold value deviated from the exact median. In other words, we considered cases where the prior information about the label distribution was vague and not precisely known. On the other hand, we utilized the prior of noise rate in determining anchors to be used in training. Based on previous findings, the average accuracy for the transfer function at test-time was approximately 80 percent, indicating a noise rate of 20 percent. In Table \ref{tab:res3}, we examined scenarios where the noise rate deviated from this established percentage.

\subsection{Data-set Size}
\label{sec:datasetsize}
\begin{figure}[hbt!]
\begingroup
    \centering
    \begin{tabular}{c}
\hspace{-0.0cm}{ \includegraphics[width=1\linewidth]{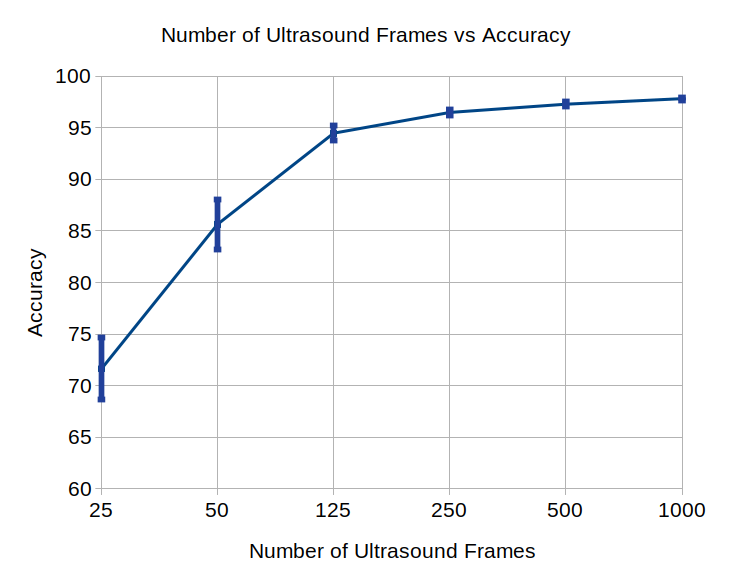}}\\
\end{tabular}
\caption{Number of Ultrasound Frames Needed for Successful Attack. Error bars represent one standard deviation}
\label{res:data_points}
\endgroup
\end{figure}

We examined the number of ultrasound frames needed from the perpetrator machines to achieve a successful attack on the victim model. In Fig. \ref{res:data_points}, on the y-axis, we depicted the classification accuracy achieved on the perpetrator machine using the proposed approach and on the x-axis, we depicted the number of ultrasound frames utilized from the perpetrator machine. In this result, we performed two iterations to refine pseudolabels, setting $\epsilon$ to 0.8.

\subsection{Different Transducers}
\label{sec:diff_trasnducers}
We examined the classification accuracy and AUC when a different transducer was used in the perpetrator machine. Specifically, the L11-5 transducer was used for the perpetrator machine, and the L9-4 transducer was used for the victim machine. Table \ref{tab:res4} reports the results for different iteration numbers. In this experiment, we set $\epsilon$ to 0.8

\begin{table}[t]
    \centering
\caption{Different Transducers}
{\renewcommand{\arraystretch}{1.2}
{\small
\begin{tabular}{ |p{1.2cm}||p{1.5cm}||p{1.8cm}|}
 \hline
\textit{Iteration}&\textit{Accuracy}&\textit{AUC}\\
 \hline
2& \textbf{96.87$\pm$1.39} & \textbf{0.9948$\pm$4e-3} \\
\hline
5& 93.55$\pm$5.00 & 0.9718$\pm$3e-2  \\
\hline
10& 88.53$\pm$5.81 & 0.9391$\pm$5e-2\\ \hline
\end{tabular}
}}
\label{tab:res4}
\end{table}
\section{Discussion}
\label{sec:discussion}

In this study, the proposed approach utilized a recently emerged transfer function within an iterative learning schema. The transfer function aids in decreasing noise in the pseudo-labeling process, while the iterative schema further refines the labels. The study's key observation highlights the ease with which the functionality of DL-based QUS methods can be stolen, raising valid concerns regarding model security. The ease with which a competitor company can transfer this functionality to its own machines poses a risk, potentially causing delays in deploying DL-based QUS methods in clinical settings. Therefore, as an outcome of the study, we emphasize the need for a deeper understanding of model security and the development of secure deployment methods tailored for ultrasound imaging contexts. Developing techniques to safeguard machine learning models emerges as a critical challenge in the adoption of wider clinical applications of machine learning-based algorithms. As of today, assuming one can access the input-output interface of a victim ultrasound machine, there is no known defense against such model extraction attacks \cite{varunspaper}. 

The study primarily aligns with the black-box UDA framework, wherein we lacked access to the internal workings of the victim model and solely utilized its input-output interface. Leveraging this interface, we obtained pseudo-labels from the unlabeled data of the perpetrator machine. Using these pseudo-labels we were able to train the DL model on the perpetrator machine without the need for annotation and labelling of the data by an outside expert. Our primary aim was to demonstrate the ease of replicating the functionality of the victim machine. While there might be technical differences such as model architecture between the victim and perpetrator machines, our focus remained on the functionality, considering it as the pivotal aspect of our investigation. 

In Table \ref{tab:res1}, we examined the impact of integrating the transfer function within the iterative schema. The findings highlighted a significant positive impact of the transfer function on the stealing performance. In the absence of the transfer function, the iterative schema yielded accuracy scores ranging between 85-90 percent. However, upon incorporating the transfer function, the performance consistently surged to the 98 percent range across all iteration numbers. Its impact on the AUC score also demonstrated a positive effect, elevating the AUC score to 0.998. This table illustrated the substantial contribution brought by the transfer function to the iterative schema. While the study focused on the utilization of the IterLNL schema, the transfer function holds the potential to similarly benefit all other black-box UDA methods. 

In Table \ref{tab:res2}, we explored the effects of the prior about the label distribution. In our experimental setup, the DL models generated a single score. When the class score approached 0, the likelihood of it being class 0 increased. When the score approached 1, the likelihood of it being class 1 increased. Following that, in Algorithm \ref{alg:algo_iter_test}, during the acquisition of noisy labels $y_{perpetrator}$ using $F$, we employed our prior knowledge of the label distribution to set the threshold between class 0 and 1. As we operated within a balanced binary classification setup, we determined the threshold as the median value ($50^{th}$ percentile). In Table \ref{tab:res2}, we investigated scenarios involving inaccurate or approximate prior knowledge of the label distribution such as setting threshold at $40^{th}$ or $60^{th}$ percentile during the acquisition of noisy labels. We observed that the proposed method demonstrated robust performance across various scenarios, maintaining an accuracy score consistently above 97 percent and an AUC score of 0.997. As a future direction, it is possible to explore various problem scenarios, including imbalanced classification, to increase the impact of the proposed approach. Moreover, the proposed approach can be scaled to a multi-class problem setting, as in certain clinical tasks, there may be multiple types of tissue states. As long as we have access to the input-output interface and obtain pseudo-labels in a multi-class setting, we expect the proposed approach to scale well.

In Table \ref{tab:res3}, we investigated the effects of the prior about the noise rate $\epsilon$. The noise rate plays a crucial role in the iterative schema, influencing the selection of anchors at each epoch based on this parameter. Given the prior work on transfer function, the noise rate was estimated at 20 percent so that 80 percent of the data was utilized as anchors during the training. In Table \ref{tab:res2}, we investigated scenarios involving inaccurate or approximate prior knowledge of the noise rate such as $10$ and $30$ percent during the acquisition of anchors. We observed that the proposed method demonstrated robust performance across various scenarios, maintaining an accuracy score consistently above 97 percent and an AUC score of 0.993.

In Fig. \ref{res:data_points}, we investigated the number of ultrasound frames needed from the perpetrator machine to achieve a successful attack. We used 25, 50, 125, 250, 500 and 1000 frames in the figure. Classification accuracies on the perpetrator were calculated by first performing patch extraction, obtaining pseudolabels, refining the labels, and training a machine learning model for the perpetrator machine. Then, using a hidden set, we obtained classification accuracies. We observed that by just utilizing 125 ultrasound frames from the perpetrator machine, the proposed approach achieved a 94.48\% classification accuracy. Developing a more data-efficient attack could be another direction for research. 

In Table \ref{tab:res4}, we investigated the effect of using different transducers between the victim and the perpetrator machines. This scenario is relevant, as different machines could use their own customized transducers. In such a scenario, the stealing algorithm needs to be robust against these variations. We achieved a 97 percent classification accuracy and a 0.9948 AUC score on the perpetrator machine. However, the performance degraded as the number of iterations increased, similar to the case when the transfer function was neglected in the approach. That could indicate that as the pseudolabels become noisier, degradation occurs when increasing the iteration number. One explanation could be that the approach is more sensitive to hyperparameters such as the iteration number or noise rate.

Although the proposed approach does not require any expert annotation, it assumes the availability of unlabeled data from the perpetrator machine. Depending on the clinical task, acquiring the unlabeled data could be as costly as acquiring expert annotation, as it requires finding the right type of patients, addressing privacy issues, and conducting lab experiments. Furthermore, as a follow-up research direction, combining data with labels acquired from the victim machine with a small subset of expert-annotated data from the perpetrator machine could be interesting and could lead to higher accuracy and more efficient attacks.  

The proposed method achieved a remarkable 98\% classification accuracy and 0.998 AUC score on the perpetrator machine by solely using the unlabeled data from the perpetrator machine and the input-output interface of the victim model. It's important to note that our assumptions included the ease with which the perpetrator machine could utilize the input-output interface, converting its data into the appropriate format and generating pseudo-labels. Furthermore, under the hood, we assumed the perpetrator has complete knowledge about the clinical task that the victim model is trained for. Therefore, the perpetrator acquires meaningful unlabeled data for the victim model. For instance, if the victim model characterizes tissue for certain organs and for certain patient distribution, we assumed the perpetrator has the complete knowledge of it. Implementing security enhancements might involve measures like hiding or restricting access to the input-output interface. Nevertheless, there remains a risk of hacking, especially in a development environment if the competitor gains ownership of the victim machine. An intriguing approach to improve model security could involve integrating security measures into the model's behavior, aiming to deceive the perpetrator machine by presenting misleading functionalities. In essence, this study underscores the critical need for an enhanced understanding of model security within the domain of DL-based ultrasound imaging. Hence, we identify model security as a critical future direction encompassing two key aspects: first, the development of new strategies to transfer DL-based functionalities to the perpetrator machine; and second, the development of security measures to defend against these potential attacks. The dataset can be accessed using the link: https://uofi.box.com/s/d9ecw002ree6gj9tlplz7t0i2f1ojbk7, and the code for this study is available at the repository:https://github.com/usoylu2/theartsteal. 
\section{Conclusion}
\label{sec:conclusion}

We demonstrated the ability to adeptly steal the functionality of a DL model from a victim machine, raising valid concerns regarding model security. The ease of transferring functionality to competitors poses a significant risk to future DL-based development in diagnostic ultrasound. The study emphasizes the need for a deeper understanding of model security and the development of secure deployment methods tailored for DL-based ultrasound imaging.

\bibliographystyle{IEEE_ECE}
\bibliography{LaTeX/tmi}

\end{document}